  [1999/12/01 v1.4c Il Nuovo Cimento]
\documentclass[varenna]{cimento}

\usepackage{graphicx}

\title{Accelerometer using atomic waves for space\\applications}

\author{A. Landragin,
F. Pereira Dos Santos}

\institute{LNE-SYRTE, CNRS UMR8630, Observatoire de Paris, UPMC\\61 avenue de l'observatoire\\ 75014 Paris, France}

\begin{document}

\maketitle

\begin{abstract}
The techniques of laser cooling combined with atom interferometry make possible the realization of very sensitive and accurate inertial sensors like gyroscopes or accelerometers. Besides earth-based developments, the use of these techniques in space should provide extremely high sensitivity for research in fundamental physics, Earth's observation and exploration of the solar system.  
\end{abstract}

\section{Introduction}
Inertial sensors are useful devices in both science and industry. Higher precision sensors could find scientific applications in the areas of general relativity~\cite{chow}, geodesy and geology. There are also important applications of such devices in the field of navigation, surveying and analysis of earth structures. Matter-wave interferometry was envisaged for its potential to be an extremely sensitive probe for inertial forces~\cite{clauser}. In 1991, atom interference techniques have been used in proof-of-principle work to measure rotations~\cite{Borde91} and accelerations~\cite{chu}. In the following years, many theoretical and experimental works have been performed to investigate this new kind of inertial sensors~\cite{atinter}. Some of the recent works have shown very promising results leading to a sensitivity comparable to other kinds of sensors, as well as for rotation~\cite{todd} as for acceleration~\cite{achim} and possibility of realizing a full inertial base within the same device~\cite{canuel}. The most developed atom-interferometer inertial sensors are today atomic state interferometers \cite{Borde:1989} which use two-photon velocity selective Raman transitions~\cite{Kas91a,Mol92} to manipulate atoms while keeping them in long-lived ground states. 

Atom interferometry is nowadays one of the most promising candidates for ultra-precise and ultra-accurate measurement of gravito-inertial forces on ground or for space~\cite{APB}. From performances on ground, one can expect unprecedented sensitivity in space, leading to many mission proposals since 2000~\cite{Hyper}.  This technology is now mature enough that several groups are developing instruments for practical experiments: in the field navigation ~\cite{applic-gradio}, or fundamental physics (gradiometer for the measurement of G~\cite{gradio}, gravimeter for the watt balance experiment~\cite{wattb}, interferometer for the measurement of fine structure constant thanks to $h/m$~\cite{hsurm}). Moreover, the realization of Bose-Einstein condensation (BEC) of a dilute gas of trapped atoms in a single quantum state~\cite{Anderson95,Davis95,Bradley95} has produced the matter-wave
analog of a laser in optics~\cite{Mewes97,Anderson98,Hagley99,Bloch99} and open new possibilities. Alike the revolution brought by lasers in optical interferometry~\cite{chow,Ste95,Stedman97},  it is expected that the use of Bose-Einstein condensed atoms will bring the science of atom optics, and in particular atom interferometry, to an unprecedented level of accuracy \cite{Bouyer:1997,Gupta:2002}. In addition, BEC-based coherent atom interferometry would reach its full potential in space-based applications where micro-gravity will allow the atomic interferometers to reach their best performance. Applications of accelerometer in space concern fundamental physics, like testing the equivalence principle by comparing the free fall of two different atomic species~\cite{mwxg}, and more generally testing all aspects of gravity as in~\cite{SAGAS}, which aims at testing the laws of gravity at large scale, as well for fundamental physics as for exploration of the solar system. 

In the following part, we investigate the sources of noise and systematic effects for such atom interferometers, thanks to the results obtained with the gravimeter under development in our laboratory. This setup is based on same concepts than previous works~\cite{achim}: atomic source realized with a magneto-optical traps and manipulation of the atomic wavepackets with Raman transitions. In our experiment, we have studied in details the influence of any perturbations on the sensitivity of the sensor. Reducing there impacts on the noise of the instrument, we finally reach an excellent sensitivity of $1.4\times10^{-8}\,\rm{m.s^{-2}.Hz^{-1/2}}$, despite a rather short interrogation time ($100 \;{\rm ms}$ only). We think this experiment is a good benchmarck to oversee the performances of best space accelerometers based on same technologies. Performances in space environment will be derived, taking into account the specifications of the environment and the much longer interrogation time.

\section{Experimental setup}
\label{expsetup}

The principle of our gravimeter is based on the coherent splitting
of matter-waves by the use of two-photon Raman transitions \cite{Kasevich91}. These
transitions couple the two hyperfine levels $F=1$ and $F=2$ of the
$^5S_{1/2}$ ground state of the $^{87}$Rb atom. An intense beam of
slow atoms is first produced by a 2D-MOT. Out of this beam $10^7$
atoms are loaded within 50 ms into a 3D-MOT and subsequently
cooled in a far detuned (-25 $\Gamma$) optical molasses. The
lasers are then switched off adiabatically to release the atoms
into free fall at a final temperature of $2.5~\mu\textrm{K}$. Both
lasers used for cooling and repumping are then detuned from the
atomic transitions by about 1~GHz to generate the two off-resonant
Raman beams. For this we have developed a compact and agile laser
system that allows us to rapidly change the operating frequencies
of these lasers, as described in \cite{Cheinet06}. Before entering
the interferometer, atoms are selected in a narrow vertical velocity
distribution ($\sigma_v \leq v_r = 5.9$ {mm/s}) in the $\left|F=1,
m_F=0\right\rangle$ state, using a combination of microwave and
optical Raman pulses.

The interferometer is created by using a sequence of three pulses
($\pi/2-\pi-\pi/2$), which split, redirect and recombine the
atomic wave packets. Thanks to the relationship between external
and internal state \cite{Borde:1989}, the interferometer phase shift
can easily be deduced from a fluorescence measurement of the
populations of each of the two states. Indeed, at the output of
the interferometer, the transition probability $P$ from one hyperfine
state to the other is given by the well-known relation for a two
wave interferometer: $P=\frac{1}{2}\left(1 +
C\cos\Delta\Phi\right)$, where $C$ is the interferometer contrast,
and $\Delta\Phi$ the difference of the atomic phases accumulated
along the two paths.  The difference in the phases accumulated
along the two paths depends on the acceleration $\vec{a}$
experienced by the atoms. It can be written as \cite{Kasevich91}
$\Delta\Phi=\phi(0)-2\phi(T)+\phi(2T)=-\vec{k}_{eff} \cdot
\vec{a}T^{2}$, where $\phi(0,T,2T)$ is the difference of the
phases of the lasers, at the location of the center of the atomic
wavepackets, for each of the three pulses \cite{Borde03}. Here
$\vec{k}_{eff}=\vec{k}_{1}-\vec{k}_{2}$ is the effective wave
vector (with $|\vec{k}_{eff}|=k_1+k_2$ for counter-propagating
beams), and $T$ is the time interval between two consecutive
pulses.

The Raman light sources are two extended cavity diode lasers based
on the design of \cite{Baillard06}, which are amplified by two
independent tapered amplifiers. Their frequency difference is
phase locked onto a low phase noise microwave reference source. The two
overlapped beams are injected in a polarization maintaining fiber,
and guided towards the vacuum chamber. We obtain
counter-propagating beams by placing a mirror and a quarterwave
plate at the bottom of the experiment. Four beams are actually
sent onto the atoms, out of which only two will drive the
counter-propagating Raman transitions, due to conservation of
angular momentum and the Doppler shift induced by the free fall of
the atoms.

Experimental setups, based on the same principle, can be realized for space experiments. They will benefit from the technical developments realized in the frame of the PHARAO atomic clock \cite{pharao} for the space ACES project.

\section{Sensitivity of the interferometer}

\subsection{Sensitivity function}

The sensitivity function characterizes the
influence of fluctuations in the Raman lasers phase difference  $\phi$ onto the
transition probability \cite{Dick87}, and thus on the
interferometer phase. This function
is defined by :

\begin{equation}
\label{eq} \ g(t)=2 \lim_{\delta \phi\rightarrow 0} \frac{\delta
P(\delta \phi,t)}{\delta \phi }.\
\end{equation}
where $\delta \phi$ is a jump on the Raman phase difference $\phi$, which occurs at time t during the
interferometer sequence, and induces a change of $\delta P(\delta
\phi,t)$ in the transition probability.

The expression of the sensitivity function can easily be derived when considering
that the Raman pulses are infinitesimally short. In that case, the
interferometer phase $\Phi$ is given by \cite{chu}:
$\Phi=\phi_1-2\phi_2+\phi_3$, where
$\phi_1$,$\phi_2$,$\phi_3$ are the the Raman laser phase differences at the three laser interactions,
taken at the position of the center of the atomic wavepacket \cite{Borde03}. Usually, the
interferometer is operated at mid fringe ($\Phi=\pi/2$), in order to maximize the sensitivity to
interferometer phase fluctuations. If the phase step $\delta
\phi$ occurs for instance between the first and the second pulses,
the interferometric phase changes by $\delta\Phi=-\delta\phi$, and
the transition probability by $\delta P=-cos(\pi/2+\delta \Phi)/2
\sim -\delta \phi/2$ in the limit of an infinitesimal phase step.
Thus, in between the first two pulses, the sensitivity function is
-1. The same way, one finds for the sensitivity function between
the last two pulses : +1.

In the more general case of finite duration Raman laser pulses, the
sensitivity function will depend on the time evolution of the atomic state
during the pulses. This function is calculated in \cite{Cheinet07} considering laser waves as plane
waves and quantizing atomic motion in the direction
parallel to the laser beams, in the case of a constant Rabi frequency (square pulses) and resonance condition fulfilled.
We calculated the change in the transition probability for a
infinitesimally small phase jump at any time t during the
interferometer, and deduce $g(t)$.

The sensitivity function is an odd function, whose
expression is given here for $t>0$:
\begin{equation}
\label{biggeq} g(t)=\left\{\begin{array}{ll}
 \sin(\Omega_R t) & 0<t<\tau_R \\
 1 & \tau_R <t<T+\tau_R\\
 -\sin(\Omega_R (T-t)) & T+\tau_R<t<T+2\tau_R\\
\end{array}\right.
\end{equation}
where we choose the time origin at the middle of the second
Raman pulse and where $\Omega_R/2\pi$ is the Rabi frequency.

Using this function, we can now evaluate the
fluctuations of the interferometric phase $\Phi$ for an arbitrary
Raman phase noise $\phi (t)$ on the lasers
\begin{equation}
\delta\Phi=\int_{-\infty}^{+\infty}g(t)d\phi(t)=\int_{-\infty}^{+\infty}g(t)\frac{d\phi(t)}{dt}dt.
\end{equation}

\subsection{Influence of the phase noise onto the sensitivity of the interferometer}

The sensitivity of the interferometer is characterized by the
Allan variance of the interferometer phase fluctuations,
$\sigma^{2}_{\Phi}(\tau)$, defined as

\begin{eqnarray}
 \sigma_{\Phi}^{2}(\tau)&=&\frac{1}{2}\langle(\bar{\delta \Phi}_{k+1}-\bar{\delta \Phi}_{k})^{2}\rangle \\
&=&\frac{1}{2}\lim_{n\rightarrow \infty}\left\{
 \frac{1}{n}\sum_{k=1}^{n}(\bar{\delta \Phi}_{k+1}-\bar{\delta \Phi}_{k})^{2}\right\}.\label{eq:variance_allan}
 \end{eqnarray}

where $\bar{\delta \Phi}_{k}$ is the average value of $\delta
\Phi$ over the interval $[t_{k},t_{k+1}]$ of duration $\tau$. The
Allan variance is equal, within a factor of two, to the variance
of the differences in the successive average values $\bar{\delta
\Phi}_{k}$ of the interferometric phase. As the interferometer is
operated sequentially at a rate $f_c=1/T_{\rm{c}}$, $\tau$ is a
multiple of $T_c$ : $\tau=m T_c$.

In order to evaluate correctly the stability of the interferometer phase $\Phi$,
it is necessary to take into account that the measurement is
pulsed. The sensitivity of the interferometer is limited by
an aliasing phenomenon, similar to the Dick effect for atomic
clocks\,\cite{Dick87}: only the phase noise at multiples of the
cycling frequency contribute to the Allan variance, weighted by the
Fourier components of the transfer function. For large averaging times $\tau$,
the Allan variance of the interferometric phase is given by
\begin{equation}
\label{Dick} \sigma^{2}_{\Phi}(\tau)={1\over
\tau}\sum_{n=1}^{\infty}|H(2\pi n f_{\rm{c}})|^2
        S_{\phi}({2\pi n f_{\rm{c}}})
        \label{eq:dick}
\end{equation}
where $S_{\phi}(\omega)$ is the power spectral density of the
Raman phase, and $H(\omega)$ is the transfer function, given by $H(\omega)=\omega G(\omega)$, where $G$ is the Fourier transform
of the sensitivity function.
\begin{equation}
G(\omega)= \int_{-\infty}^{+\infty}e^{-i\omega t}g(t)dt
\label{eq:G1}
\end{equation}

At low frequency ($\omega<<\Omega_R$), $G$ can be approximated by
\begin{equation}
G(\omega)=-\frac{4i}{\omega}\sin^2[\omega  (T+2\tau_R)/2] \label{eq:G4}
\end{equation}
The transfer function $|H^{2}|$ has two important
features. First, it presents oscillations at a frequency
given by $1/(T+2\tau_R)$, leading to zeros at frequencies given by
$f_k=\frac{k}{T+2\tau_R}$. The second is a low pass first order
filtering due to the finite duration of the Raman pulses.

For white phase noise $S_{\phi}^0$, and to first order in
$\tau_R/T$, the phase stability is given by:
\begin{equation}
\label{whiteeq1} \sigma^{2}_{\Phi}(\tau)=\frac{\pi \Omega}{2}
S_{\phi}^0\frac{T_c}{\tau}
\end{equation}
where $\Omega$ is the Rabi frequency.

This illustrates the filtering of the transfer function: the shorter
the pulse duration $\tau_R$, the greater the interferometer noise.

\subsection{The 100 MHz source oscillator}
\label{sec:1}

To reduce the noise of the interferometer, the frequency
difference between the Raman beams needs to be locked to a very
stable microwave oscillator, whose frequency is close to the
hyperfine transition frequency ($\nu_{mw}=6.834$ GHz for
$^{87}$Rb). The reference frequency will be
delivered by a frequency chain, which transposes in the microwave
domain the low phase noise of an RF oscillator, typically a quartz
oscillator. When this transposition is performed without
degradation, the phase noise power spectral density of the RF
oscillator, of frequency $\nu_{rf}$, is multiplied by
$(\nu_{mw}/\nu_{rf})^2$.

\begin{figure}[h]
  \begin{center}
\includegraphics{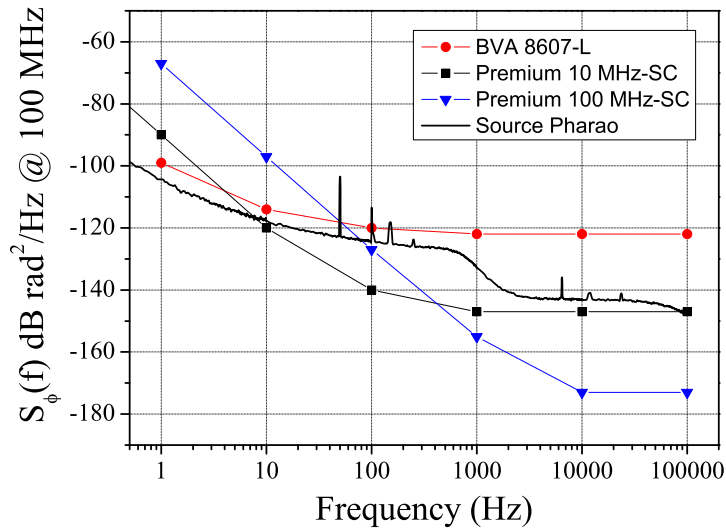}
 \caption{Specifications for the phase noise spectral density of
different quartz oscillators, transposed at 100 MHz. The phase
noise of the source developed for the Pharao project is also
displayed as a solid black line.\label{fig:specquartz}}
 \end{center}
\end{figure}

No single quartz oscillator fulfills the requirements of ultra low
phase noise over a sufficiently large frequency range. We present
in figure \ref{fig:specquartz} the specifications for different
high stability quartz : a Premium 10 MHz-SC from Wenzel, a BVA
OCXO 8607-L from Oscilloquartz, and a Premium 100 MHz-SC quartz
from Wenzel. The phase noise spectral density is displayed at the
frequency of 100 MHz, so that one can compare the different quartz
oscillators.

A 100 MHz source for a space interferometer could be realized by
combining two quartz oscillators. A 100 MHz quartz would be locked
onto one of the above mentioned high stability 10 MHz reference
oscillators. The bandwidth of this lock will correspond to the
frequency below which the phase noise of the reference quartz is
lower than the noise of the 100 MHz quartz.

The phase noise properties of such a combined source can be seen
in figure \ref{fig:specquartz}, where we display as a solid
line the phase noise spectral density of the 100 MHz source
developed by THALES for the PHARAO space clock project. This
combined source has been optimized for mimimal phase noise at low
frequency, where it reaches a level of noise lower than any
commercially available quartz. An atomic clock is indeed mostly
limited by the lower frequency part of the frequency spectrum, so
the requirements on the level of phase noise at higher frequency
($f>1$kHz) is less stringent than for an atom interferometer. A
medium performance 100 MHz oscillator is thus sufficient.
 
Using a simple model for the phase lock loop, we calculated the
phase noise spectral density of the different combined sources we
can realize by locking the Premium 100 MHz-SC either on the
Premium 10 MHz-SC (Source 1), or on the BVA (source 2), or even on
the Pharao source (source 3). We then estimated the impact on the
interferometer of the phase noise of the 100 MHz source, assuming
we are able to transpose the performance of the source at 6.8 GHz
without degradation. We calculated using \ref{eq:dick} the Allan
standard deviation of the interferometric phase fluctuations for
the different configurations and for various interferometer
parameters. The results are presented in table \ref{tb:noise}.

\begin{table}[h]
    \centering
            \begin{tabular}{|l|c|c|c|c|c|}
                    \hline

                                &            &             & Source 1 & Source 2 & Source 3\\
                    {\bf $T_c$} & {\bf $2T$} & {\bf $\tau_R$} & {\bf $\sigma_{\Phi}$($T_c$)} & {\bf $\sigma_{\Phi}$($T_c$)} &  {\bf $\sigma_{\Phi}$($T_c$)}\\
                    {\bf (s)} & {\bf (s)} & {\bf ($\mu$s)} & {\bf (mrad)} & {\bf (mrad)} &  {\bf (mrad)}\\

                    \hline
                            0.25  & 0.1 & 10 & 1.2 & 3.5 & 2.2 \\
                    \hline
                    3  & 2 & 10 & 21 & 6.5 & 4.4 \\
                     \hline
                    15 & 10 & 10  & 110 & 37 & 19 \\

                    \hline

             \end{tabular}
        \caption{Contribution of the 100 MHz source phase noise to the interferometrer phase
fluctuations. The calculation has been performed for a Rb interferometer,
for each of the three different sources and for various parameters
of the interferometer. First case corresponds to the usual parameters of our gravimeter, and second (resp. third) case corresponds to typical values for space experiments with a MOT atomic source (resp. ultra-cold source).}
        \label{tb:noise}
        \end{table}

For short interrogation times (such as $2T=100$ ms, which is the
maximum interrogation time of our cold gravimeter), Source 1 behaves better, whereas
for large interrogation times, where the dominant contribution to the
noise comes from the lower decades (0.1-10 Hz), Source 2 and 3 are
better.
We assumed here that for any of these sources, the phase noise below 1 Hz is
well described as flicker noise, for which the spectral
density scales as $S_{\phi}(f)= S_{\phi}(1\rm{Hz})/f^3$. If the
phase noise would behave as pure flicker noise over the whole
frequency spectrum, the Allan standard deviation of the
interferometer phase would scale as $T$. This is roughly the
behavior we notice in the table for long interrogation times.

The sensitivity of the interferometer for acceleration scales as
$T^2$, so that the sensitivity to acceleration gets better when
the interrogation time gets larger. For example, for $2T= 10 $s and
$T_c=15$ s (resp. $2T=2$ s and
$T_c=3$ s), the phase noise of source 3 would limit the
sensitivity to acceleration of the interferometer to
$1.8\times10^{-10}\rm{m.s}^{-2}$ (resp. $4.7\times10^{-10}\rm{m.s}^{-2}$) at 1 s for $^{87}Rb$.

\subsection{The frequency chain}
\label{sec:3}

The microwave signal is generated by multiplication of the 100 MHz
source. An example of the generation of the microwave reference can be found in \cite{Nyman06}.
The contribution to the interferometer noise of this system was found to be 0.6 mrad per shot for $\tau_R=10 \mu$ s, $2T=10$ s and $T_c=15$ s, which
is negligible with respect to the contribution from the 100 MHz
source.

\subsection{Propagation in the fiber}

In most of the experiment, the Raman lasers are generated by two
independent laser sources. The two beams are mixed using a polarizing beam splitter cube, so that they have
orthogonal polarizations. A small fraction of the total power can be
sent to one of the two exit ports of the cube, where a fast
photodetector detects the beat frequency, in order to phase lock the lasers. Both beams are finally
guided towards the atoms with a polarization maintaining fiber.
Since the Raman beams have orthogonal polarization, fiber
length fluctuations induce phase fluctuations, due to the
birefringence of the medium. The phase noise induced
by the propagation in the fiber can be measured by comparing the beat signal
measured after the fiber with the one we use for the phase lock.

\begin{figure}[h]
  \begin{center}
\includegraphics{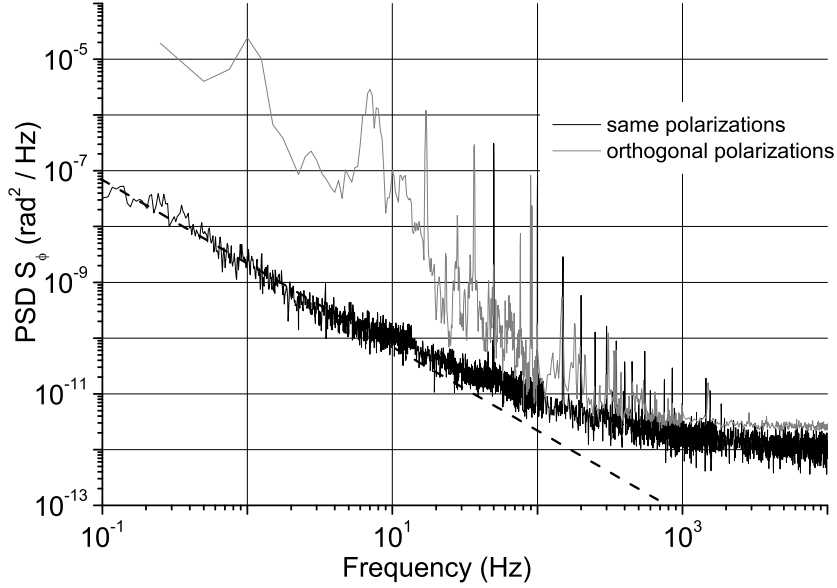}
 \caption{Contribution of the propagation in the optical fiber to the Raman phase noise. Grey curve displays the noise in the case of a polarization maintaining fiber and two orthogonal polarizations for the two Raman lasers. Black curve displays the noise in the case of a polarizing fiber and two parallel polarizations for the two Raman lasers.\label{DSPFibre}}
 \end{center}
\end{figure}

Figure \ref{DSPFibre} displays the power spectral density of the
phase noise induced by the propagation, which is dominated by low
frequency noise due to acoustic noise and thermal fluctuations.
This source of noise can be reduced by shielding the fiber from the
air flow of the air conditioning, surrounding it with some
packaging foam. 
An alternative technique consists in using identical linear polarizations for the Raman beams. 
This can be achieved using a polarizer after the mixing, by generating the Raman lasers by phase modulatuion of a single laser, or by injecting with two independent lasers a power amplifier. In this case, the noise is efficiently suppressed, as shown in figure \ref{DSPFibre}, down to a level where it is negligible.

\subsection{Detection noise}
 
Quantum projection noise limit constitutes the intrinsic limit of sensitivity and scales as $1/\surd{N}$, where N is the number of detected atoms per shot. Typical values are below 1 mrad for MOT sources and 3 mrad for ultra cold atomic sources.  Thus for long interrogation times, as for space applications, the dominant source of phase noise is expected to be due to the stability of the 100 MHz source. 

\section{The case of parasitic vibrations\label{sec:vibrations}}

The same formalism can be used to
evaluate the degradation of the sensitivity to inertial forces
caused by vibrations of the retroreflecting mirror.

The sensitivity of the interferometer is then given by
\begin{equation}
\label{Vibeq} \sigma^{2}_{\Phi}(\tau)={k_{eff}^2\over
\tau}\sum_{n=1}^{\infty}|H(2\pi n f_{\rm{c}})|^2
        S_{z}(2\pi n f_{\rm{c}})
\end{equation}
where $S_{z}(\omega)$ is the power spectral density of position
noise. Introducing the power spectral density of acceleration
noise $S_{a}(\omega)$, the previous equation can be written

\begin{equation}
\label{Vibeq2} \sigma^{2}_{\Phi}(\tau)={k_{eff}^2\over
\tau}\sum_{n=1}^{\infty}
        \frac{|H(2\pi n f_{\rm{c}})|^2}{(2\pi n f_{\rm{c}})^4}
        S_{a}(2\pi n f_{\rm{c}})
\end{equation}
It is important to note here that the acceleration noise is
efficiently filtered by the transfer function for acceleration, which
decreases as $1/f^4$.

In the case of white acceleration noise $S_{a}$, and to first
order in $\tau_R/T$, the limit on the sensitivity of the
interferometer is given by :

\begin{equation}
\label{whiteeq} \sigma^{2}_{\Phi}(\tau)=\frac{k_{eff}^2
T^4}{2}\left(\frac{2T_{\rm{c}}}{3T}-1\right)\frac{S_{a}}{\tau}\
\end{equation}

To put this into numbers, we now calculate the requirements on the
acceleration noise of the retroreflecting mirror in order to reach
a sensitivity of $\times10^{-10}\,\rm{m.s^{-2}}$ at 1 s. For negligible dead time ($Tc\simeq2T)$,
the amplitude noise should lie below $2.5\times10^{-10}\,\rm{m.s^{-2}.Hz^{-1/2}}$.

\section{Systematic effects}

Regarding the targeted ultra-high accuracies for atom inertial
sensors, many environmental parameters have to be
controlled with great precision. An important tool, that
allows us to suppress many of the remaining systematic phase
shifts, relies on the fact that we can distinguish two classes
of systematic effects. 

Phase shifts that are independent of the direction
of $\overrightarrow{k_{eff}}$ and others that change sign when
reversing $\overrightarrow{k_{eff}}$. This allows us to separate
and distinguish their influence on the atomic signal in constantly
repeated $k_\uparrow - k_\downarrow$ measurements. Assuming that
the trajectories of the atoms remain constant, the half difference
$\frac{1}{2}\triangle_{\uparrow \downarrow}$ of the measured
interferometer phase $\Delta\Phi_{int}$ of a consecutive
$k_\uparrow - k_\downarrow$ measurement will contain only phase
terms depending on $\overrightarrow{k_{eff}}$. Whereas
$\overrightarrow{k_{eff}}$ independent phase terms are contained
in the half sum $\frac{1}{2}{\sum}_{\uparrow \downarrow}$:

\begin{equation}
\frac{1}{2}\triangle_{\uparrow \downarrow}= - \cos \theta
|\overrightarrow{k}_{eff}| \cdot |\overrightarrow{g}| \cdot
T^2+\Delta \Phi_{LS2}+ \Delta \Phi_{Coriolis}+\Delta
\Phi_{Aberr}+...
 \label{eq2}
\end{equation}

\begin{equation}
\frac{1}{2}{\sum}_{\uparrow \downarrow}=\Delta \Phi_{RF}+\Delta
\Phi_{LS1}+\Delta \Phi_{gradB}+...
 \label{eq1}
\end{equation}

\subsection{k-independent Phase Shifts}

Three major contributions to large phase shifts in the
interferometer, that can be rejected from the $k$-dependent
acceleration signal, $-\overrightarrow{k_{eff}} \cdot
\overrightarrow{g} \, {T}^2$, arise from the presence of magnetic
field gradients, 1-photon light shifts \cite{weiss1994} and RF-phase shifts (see
equ.~(\ref{eq1})).

We have experimentally demonstrated this rejection on our cold atom gravimeter.
To test the rejection of k-independent phase shifts from the
actual atomic signal, we have performed differential measurements with 4
configurations (two sets of parameters and $k_\uparrow -
k_\downarrow$ each). The difference in their half-sums $\Delta
(\frac{1}{2}\sum)$ gives us the additionally introduced phase
shift. The difference in their half-differences $\Delta
(\frac{1}{2}\triangle)$ shows the remaining phase error, that is
not rejected by the $k_\uparrow - k_\downarrow$ measurement. 
 
With this method, we demonstrated rejection efficiencies typically better than 99\%, limited by the resolution of the measurement (in the case of the RF and 1-photon light shift), and by the imperfect overlap between the atomic trajectories for the $k_\uparrow - k_\downarrow$ interferometers (in the case of the magnetic field gradient).
For a space interferometer, the influence of the magnetic field gradients are expected to be reduced with respect to the case of ground experiment, as i) atoms travel along shorter distances, and ii) the two separated wave-packets experience the same phase shifts, as they alternately travel along identical trajectories.

 \begin{figure}[h]
  \begin{center}
\includegraphics[angle=270,width = 13.5 cm]{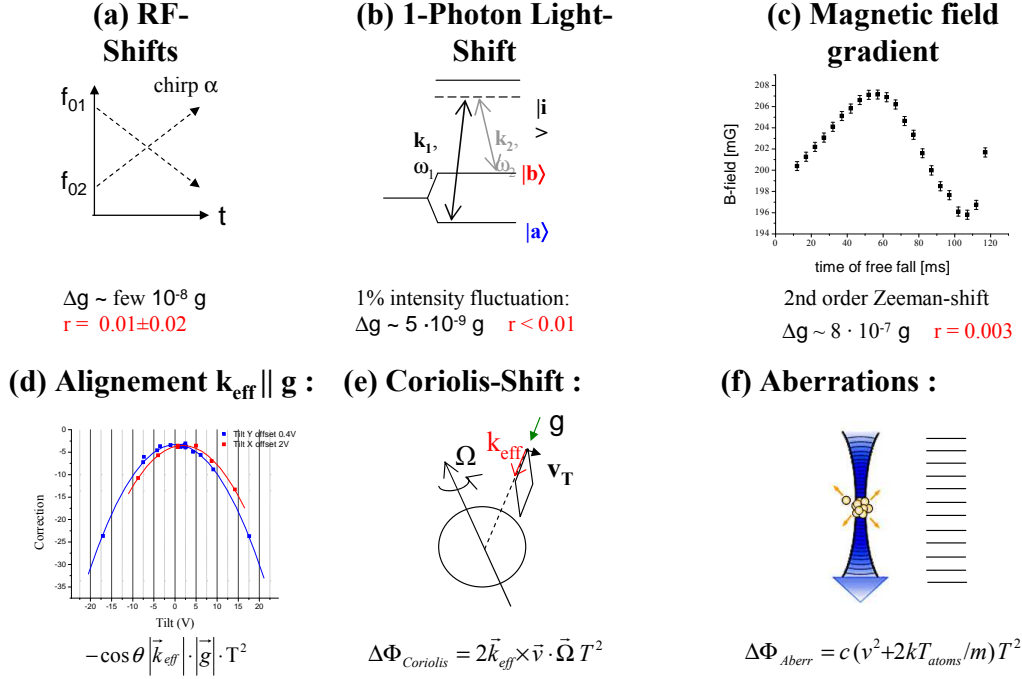}
 \caption{Some major contributions to systematic phase shifts in our cold atom gravimeter. (a)-(c) depend on the direction of $k_{eff}$ and can be rejected to a residual $r$, (d)-(e) are indistinguishable from the acceleration signal and need to be controlled within a high precision.\label{fig:all} }
 \end{center}
\end{figure}

\subsection{k-dependent Phase Shifts}

As shown above, large systematic error contributions can be
removed from the atomic signal by systematic
$k_{\uparrow}$-$k_{\downarrow}$ measurements. The remaining phase
shifts in equation~(\ref{eq2}) are due to effects that are
inherently sensitive to the direction of $k$. 

The first term in equ.~(\ref{eq2}) represents the actual atomic
signal due to the acceleration of the atoms. The atomic
transition wavelength, which is determining $k_{eff}$, is known
to better than 1 kHz~\cite{ye1996}. 

Three major sources of phase errors add onto this acceleration
induced phase shift. Similar to the 1-photon light
shift, the Eigen-energies of the atomic states are modified in
higher order terms by 2-photon transitions of the Raman beams
themselves (LS2). This 2-photon light shift introduces a phase
shift similar to this of the usual 1-photon light shift, but where the shift of the resonance transition is 
${\delta\nu_{AC}}^{(LS2)}={\Omega_{eff}}^2/\delta_{Doppler}$, and
$\delta_{Doppler}=k_{eff} \cdot v$ is the Doppler-detuning of the
stimulated Raman transition from resonance.

Term 3 and 4 in equ.(\ref{eq2}) are phase terms that both depend
on the transverse velocities of the atoms and are difficult to
clearly distinguish. Any residual offset velocity $v_T$
perpendicular to $\overrightarrow{k_{eff}}$ will lead to phase
errors due to Coriolis forces and imperfect plane wave fronts of
the Raman beams (aberrations) \cite{Fils05}. As indicated in
figure~\ref{fig:all}(e), a finite $v_T$ will lead to a finite area
spanned by the interferometer and thus will make it sensitive to
rotations. 

Contrary to that, phase shifts due to aberrations in the phase
fronts of the Raman beams scale with the temperature $T_{atoms}$
of the atomic ensemble, depending on the shape of distortion in
the wave fronts. For a in first order parabolic curvature of the
wave front $\Delta \Phi=c \cdot r^2$, we obtain
$\Delta\Phi_{Aberr}=c \,(v_T^2+2\,k_B\,T_{atoms}/m)T^2$. Here, the
interferometer is only sensitive to perturbations introduced in
the retro-reflected beam path, which is non-common to both Raman
laser beams. In case of a pure curvature of the retro-reflected
phase fronts, $\Delta a=10^{-10}$ m/s$^2$ would require $c=17$ rad/m$^2$, which
corresponds to a phase front radius of $R=240$ km, or a flatness
of $\approx \lambda/1500$ over a beam diameter of 10 mm. This flatness is difficult to reach, especially with retardation plates. 

Irregularly shaped phase distortions, as introduced by mirror and
$\lambda/4$ plate, can average out and the requirement for the
surface flatness is less stringent. To deduce the effect on the
atomic signal, the wave front distortion introduced by the individual optical elements have to be measured with a
Shack-Hartmann sensor or ZYGO interferometer, at the level of $\approx \lambda/1000$ or better.  The collected phase shifts
along the classical trajectories of the atoms can then be calculated. Measurements of the phase shift of the interferometer versus initial position and velocity, and versus temperatures can be useful to study this systematic effect, and correct for it.

In space, Coriolis acceleration could also be an issue, depending on the details of the mission. One should though keep in mind that this bias is not intrinsic to the device, but is a part of the signal and is related to the actual trajectory of the satellite. Wavefront aberrations appear as a very important systematic effect, which depends on a non trivial way with the experimental parameters (size and temperature of the atomic cloud and interaction time). To overcome this problem, extremely high flatness optics are required.

\section{Conclusion}
Thanks to a careful study and optimization of all sources of noise and systematics on our gravimeter, we can evaluate the sensitivity and the accuracy for a space accelerometer based on classical cold atom sources and stimulated Raman transition to manipulate the atomic wavepackets. As proposed in \cite{SAGAS}, such a accelerometer might be developed easily  taking advantage of the similarities with the atomic clock prototype PHARAO for which key components have already been realized and tested on ground for the Engineer Model \cite{pharao}. Short term sensitivity should not be better than state of the art classic proof mass accelerometers \cite{onera}, but an atom interferometer should reach a much better long term stability and accuracy, without the need of spinning the satellite. Moreover, the main sources of noise and systematics cancel in a differential measurement, as in the cases of gradiometers \cite{maleki} or gyroscopes. 

Finally, the use of ultra-cold atoms (like BEC) will take full advantage of the space environment, by allowing to increase the measurement time and reducing the systematics (the effect of wavefront distortion for example) \cite{mwxg}. Such experiments are more complicated and still need to demonstrate their possibilities on ground. This is why several projects are currently developed to improve the knowledge and technology for zero-g environment: the QUANTUS project \cite{quantus} carry on in Germany, which study the realization of BEC in the free falling Bremen tower, the Frech project ICE \cite{Nyman06}, which tests the realization of an atom interferometer in the zero-g airplane of CNES, or the ESA project "Atom Interferometry Sensors for Space Applications". All these developments will soon give the possibilitiy of using such a device in fundamental physics experiments or in other applications in the field of Earth observation, navigation or exploration of the solar system.

\acknowledgments
We would like to thank all our collaborators of the "Inertial Sensors Team" of SYRTE laboratory  for their contributions to the experimental results. We also supports the Institut Francilien pour la Recherche sur les Atomes Froids (IFRAF), the European
Union (FINAQS), the D\'{e}l\'{e}gation G\'{e}n\'{e}rale pour l'Arrmement and the Centre Nationale d'Etudes Spatiales for their financial supports.

\end{document}